\documentclass[twoside]{article}
\usepackage{amsmath}
\usepackage[psamsfonts]{amssymb}
\usepackage{cmmib57}
\usepackage{diagrams}
\newcommand{\bPf}{\par\vspace*{-4pt}\indent{\sc Proof.}\enskip}
\newcommand{\ePf}{\medskip}
\def\QED{\hskip0.1em\hfill\null\ \null\nobreak\hfill\kern3pt\vbox{\hrule\hbox
    {\vrule\kern1pt\vbox{\kern1.7pt\hbox{$\scriptscriptstyle{QED}$}
     \kern0.2pt}\kern1pt\vrule}\hrule}}

\def\END{\hskip0.1em\hfill\null\ \null\nobreak\hfill\kern3pt\vbox{\hrule\hbox
    {\vrule\kern1pt\vbox{\kern1.7pt\hbox{$\,\,\,\vspace{5pt}$}
     \kern0.2pt}\kern1pt\vrule}\hrule}}
\newtheorem{theorem}{Theorem}
\newtheorem{lemma}{Lemma}
\newtheorem{corollary}{Corollary}
\newtheorem{proposition}{Proposition}
\newtheorem{remark}{Remark}
\newtheorem{definition}{Definition}
\newtheorem{example}{Example}
\DeclareMathOperator{\Diff}{Diff}
\newcommand{\bEq}{\begin{eqnarray}}
\newcommand{\eEq}{\end{eqnarray}}
\newcommand{\beq}{\begin{eqnarray*}}
\newcommand{\eeq}{\end{eqnarray*}}
\newcommand{\bDf}{\begin{definition}\em}
\newcommand{\eDf}{\end{definition}}
\newcommand{\bLm}{\begin{lemma}}
\newcommand{\eLm}{\end{lemma}}
\newcommand{\bPr}{\begin{proposition}}
\newcommand{\ePr}{\end{proposition}}
\newcommand{\bTh}{\begin{theorem}}
\newcommand{\eTh}{\end{theorem}}
\newcommand{\bCr}{\begin{corollary}}
\newcommand{\eCr}{\end{corollary}}
\newcommand{\bRm}{\begin{remark}\em}
\newcommand{\eRm}{\end{remark}}
\newcommand{\bEx}{\begin{example}\em}
\newcommand{\eEx}{\end{example}}

\newcommand{\ie}{{\em i.e$.$} }
\newcommand{\eg}{{\em e.g$.$} }
\newcommand{\R}{I\!\!R}

\newcommand{\mto}{\mapsto}

\newcommand{\ucar}[1]{\underset{#1}{\times}}

\newcommand{\owed}[1]{\overset{#1}{\wedge}}

\newcommand{\cA}{\mathcal{A}}
\newcommand{\cC}{\mathcal{C}}

\newcommand{\cE}{\mathcal{E}}
\newcommand{\cH}{\mathcal{H}}
\newcommand{\cJ}{\mathcal{J}}

\newcommand{\cL}{\mathcal{L}}

\newcommand{\by}{\boldsymbol{y}}

\newcommand{\bF}{\boldsymbol{F}}

\newcommand{\bG}{\boldsymbol{G}}

\newcommand{\bP}{\boldsymbol{P}}

\newcommand{\bX}{\boldsymbol{X}}
\newcommand{\bY}{\boldsymbol{Y}}
\newcommand{\bW}{\boldsymbol{W}}
\newcommand{\bZ}{\boldsymbol{Z}}

\newcommand{\sub}{\subset}

\newcommand{\wed}{\wedge}
\newcommand{\com}{\!\circ\!}
\newcommand{\ten}{\!\otimes\!}
\newcommand{\alp}{\alpha}
\newcommand{\bet}{\beta}
\newcommand{\gam}{\gamma}
\newcommand{\del}{\delta}
\newcommand{\eps}{\epsilon}
\newcommand{\zet}{\zeta}

\newcommand{\lam}{\lambda}

\newcommand{\ome}{\omega}

\newcommand{\Lam}{\Lambda}

\newcommand{\Ome}{\Omega}

\newcommand{\vartht}{\vartheta}

\newcommand{\For}{{\Lambda}}
\newcommand{\Con}{{\mathcal{C}}}
\newcommand{\Hor}{{\mathcal{H}}}
\newcommand{\Var}{{\mathcal{V}}}
\newcommand{\Thd}{{\Theta}}
\title{
Lagrangian reductive structures \\ on gauge-natural bundles}
\author{{\normalsize M.
Palese}
\\{\footnotesize Department of Mathematics,
University of Torino}
\\{\footnotesize via C. Alberto 10, 10123 Torino, Italy}\\  {\normalsize and}\\ 
{\normalsize E. Winterroth}\\
{\footnotesize Eduard \v{C}ech Center for Algebra and Geometry}\\
{\footnotesize University of Brno, Czech Republic}\\ 
{\footnotesize e--mails: 
{\sc marcella.palese@unito.it, ekkehart.winterroth@unito.it}}}
\date{}
\begin{document}

\maketitle

\begin{abstract}
A reductive structure is associated here with Lagrangian canonically defined  conserved quantities on gauge-natural bundles. 

\noindent Infinitesimal parametrized transformations defined by the gauge-natural lift of infinitesimal principal automorphisms induce a variational sequence such that the generalized Jacobi morphism is naturally self-adjoint. As a consequence, its kernel defines a reductive split structure on the relevant underlying principal bundle.

\medskip


\end{abstract}

\section{Introduction}

Since jet spaces and formal derivatives form a natural geometric framework for the representation of partial differential equations, various differential-geometric formulations of calculus of variations on jet spaces have been proposed
(see, \eg \cite{AnPo95,GoSt73,KMS93,Kru73,Kru90,Pal68,Sau89,Tak79,Tul77,Vin77}).
In most of these formulations the differential operator transforming Lagrangians to
Euler--Lagrange expressions is nothing but a sheaf morphism of a certain differential sheaf sequence, thus providing two related frames \cite{Vit07}:  infinite order variational bicomplexes and Krupka's finite order variational sequences.
We work within the framework of {\em finite order variational sequences} on those very important geometric constructions called {\em gauge-natural bundles} \cite{Ec81,KMS93} by considering
variational derivatives of gauge-natural invariant Lagrangians of arbitrary order in the general case of $n$
independent variables and $m$ unknown functions.
As well known, following Noether's theory \cite{Noe18}, from invariance properties of the Lagrangian the existence of suitable 
conserved currents and identities can be deduced.
Within such a picture  {\em generalized Bergmann--Bianchi identities} \cite{Ber49}  are conditions for a Noether 
conserved current to be not only closed but also the global divergence of
a tensor density called a superpotential. 
First in \cite{PaWi03} and then in a series of papers \cite{FPW04,FPW05,PaWi04,PaWi07,PaWi08,Win07} we proposed an
approach to deal with the problem of {\em canonical} covariance and uniqueness of conserved quantities which uses {\em variational
derivatives} taken with respect to the class of (generalized) variation vector fields being Lie derivatives of
sections of bundles by
gauge-natural lifts of infinitesimal principal automorphisms. 
Such variational derivatives can be suitably interpreted as vertical differentials \cite{FPV02}.

In their stemming paper on the Hamilton--Cartan formalism \cite{GoSt73}, under certain assumptions on admissible variations, Goldschmidt and Sternberg found that the Jacobi morphism is self-adjoint along solutions of the Euler--Lagrange equations in first order Lagrangian field theory. Within the geometric framework of jets of fibered manifolds they proved that the Hessian morphism, which is nothing but the second variation of the action integral of a Lagrangian, is in fact a symmetric bilinear morphism. Their proof is based on the fact that variations are choosen to be vanishing on the boundary of the integration manifold, so that integrals of divergences vanish on the boundary by virtue of Stoke's theorem.  The Hessian morphism is in fact symmetric up to a term wich is the integral of a total divergence  and vanishes following standard arguments in calculus of variations. 
As an immediate consequence of the symmetry properties of the Hessian, the latter being the integral of the contraction of  the Jacobi morphism with a variation vector field, the Jacobi morphism itself is self-adjoint. This result was used for an important application of the Morse index Theorem \cite{Smale65}.

In \cite{FPV02} a {\em quotient second variational derivative}, generalizing invariantly the classical Hessian morphism  {\em  up to horizontal differentials}, was described as the vertical differential of the Euler--Lagrange morphism generalizing the classical Jacobi morphism which thus  turns out to be self-adjoint along solutions of the Euler-Lagrange equations. 

In this paper, we use intrinsic linearity properties of the gauge-natural lift functor to first prove a Lemma stating in full generality the self-adjointness of the generalized {\em gauge-natural} Jacobi morphism defined as a generalized Euler--Lagrange type morphism in a finite order variational sequence on an extended space. As one of the relevant consequence of this Lemma, we prove that the kernel of the Jacobi morphism defines a split structure on the relevant underlying principal bundle and that such a structure is also reductive. 

It is now remarkable that we are framing our investigations within a formulation of the calculus of variations on fibered manifolds, the variational sequence, which is completely differential and free from the use of integrals such as the integral of action. Variational objects such as Euler--Lagrange equations -- and thus all higher degree generalizations such as Bergamnn-Bianchi identities, generalized Noether identities, generalized Jacobi equations --  are obtained as quotient morphisms of  the exterior differential operator acting on sheaves of differential forms. In this case the integration by parts of the action is substituted by global decomposition formulae of well characterized vertical morphism (see \eg \cite{Kol83,Tul77,Vin77} and the review in \cite{Vit07}).  The problem of dealing with local divergences can be solved by using the intrinsic properties of the variational sequence itself and the very nature of variation vector fields we choose (vertical parts of gauge-natural lifts).

\section{Jets of gauge-natural bundles and Euler--\\Lagrange type morphisms}\label{1}

We recall some basic definitions and results from the
theory of jet spaces with the main aim of stating the notation.
Let $\pi : \bY \to \bX$ be a fibered manifold,
with $\dim \bX = n$ and $\dim \bY = n+m$.
For $s \geq q \geq 0$ integers we deal with the $s$--jet space $J_s\bY$ of 
$s$--jet prolongations of (local) sections
of $\pi$; in particular, we set $J_0\bY \equiv \bY$. We recall that there are the natural fiberings
$\pi^s_q: J_s\bY \to J_q\bY$, $s \geq q$, $\pi^s: J_s\bY \to \bX$, and,
among these, the {\em affine\/} fiberings $\pi^{s}_{s-1}$.
We denote by $V\bY$ the vector subbundle of the tangent
bundle $T\bY$ of vectors on $\bY$ which are vertical with respect
to the fibering $\pi$.

For $s\geq 1$, we consider the following natural splitting induced by the natural contact structure on
jets bundles  (see \eg \cite{Vit98}):
\bEq
\label{jet connection}
J_{s}\bY\ucar{J_{s-1}\bY}T^*J_{s-1}\bY =\left(
J_s\bY\ucar{J_{s-1}\bY}T^*\bX\right) \oplus\cC^{*}_{s-1}[\bY]\,,
\eEq
where $\cC^{*}_{s-1}[\bY] \doteq \textstyle{Im} \vartht_s^*$ and
$\vartht_s^* : J_s\bY \ucar{J_{s-1}\bY} V^*J_{s-1}\bY \to
J_s\bY \ucar{J_{s-1}\bY} T^*J_{s-1}\bY \,$.
 
A vector field $\xi$ on $\bY$ is said to be \emph{vertical} if it takes values
in $V\bY$. A vertical vector field can be prolonged to a
vertical vector field $j_{s}\xi\colon J_s\bY\to VJ_s\bY$. The vector field
$j_s\xi$ is characterized by the fact that its flow is the natural prolongation
of the flow of $\xi$. 
Given a vector field $\Xi : J_{s}\bY \to TJ_{s}\bY$, the splitting
\eqref{jet connection} yields $\Xi \, \com \, \pi^{s+1}_{s} = \Xi_{H} + \Xi_{V}$. We shall call
$\Xi_{H}$ and $\Xi_{V}$ the  horizontal and the vertical part of $\Xi$, respectively.
As well known, the above splitting induces also a decomposition of the exterior differential
on $\bY$, $(\pi^{r+1}_r)^*\circ d = d_H + d_V$, where $d_H$ and $d_V$ are
called the \emph{horizontal} and \emph{vertical differential}, respectively.
We must stress that such decompositions always rise the order of the objects.

\medskip

Let $\bP\to\bX$ be a principal bundle with structure group $\bG$.
Let $r\leq k$ be integers and $\bW^{(r,k)}\bP$ $\doteq$ $J_{r}\bP\ucar{\bX}L_{k}(\bX)$, 
where $L_{k}(\bX)$ is the bundle of $k$--frames 
in $\bX$ \cite{Ec81,KMS93}, $\bW^{(r,k)}_{n}\bG \doteq \bG^{r}_{n}\odot GL_{k}(n)$
the {\em semidirect} product with respect to the action of $GL_{k}(n)$ 
on $\bG^{r}_{n}$ given by  
jet composition and $GL_{k}(n)$ is the group of $k$--frames 
in $\R^{n}$. Here we denote by $\bG^{r}_{n}$ the space of $(r,n)$-velocities on $\bG$ \cite{KMS93}.
The bundle $\bW^{(r,k)}\bP$ is a principal bundle over $\bX$ with structure group
$\bW^{(r,k)}_{n}\bG$.
Let $\bF$ be a manifold and $\zet: \bW^{(r,k)}_{n}\bG\ucar{}\bF\to\bF$ be 
a left action of $\bW^{(r,k)}_{n}\bG$ on $\bF$. There is a naturally defined 
right action of $\bW^{(r,k)}_{n}\bG$ on $\bW^{(r,k)}\bP \times \bF$ so that
 we have in the standard way the associated {\em gauge-natural bundle} of order 
$(r,k)$: $\bY_{\zet} \doteq \bW^{(r,k)}\bP\times_{\zet}\bF$.
All our considerations shall refer to $\bY$ as a gauge-natural bundle as just defined.

Denote now by $\cA^{(r,k)}$ the sheaf of right invariant vector fields 
on $\bW^{(r,k)}\bP$. A functorial map $\mathfrak{G}$ is defined 
which lifts any right--invariant local automorphism $(\Phi,\phi)$ of the 
principal bundle $W^{(r,k)}\bP$ to a unique local automorphism 
$(\Phi_{\zet},\phi)$ of the associated bundle $\bY_{\zet}$. 
Its infinitesimal version defines the {\em gauge-natural lift} in the 
following way:
\bEq
\mathfrak{G} : \bY_{\zet} \ucar{\bX} \cA^{(r,k)} \to T\bY_{\zet} \,:
(\by,\bar{\Xi}) \mto \hat{\Xi} (\by) \,,
\eEq
where, for any $\by \in \bY_{\zet}$, one sets: $\hat{\Xi}(\by)=
\frac{d}{dt} [(\Phi_{\zet \,t})(\by)]_{t=0}$,
and $\Phi_{\zet \,t}$ denotes the (local) flow corresponding to the 
gauge-natural lift of $\Phi_{t}$. Such a functor defines a class of parametrized contact transformations.

This mapping fulfils the following properties (see \cite{KMS93}):
 $\mathfrak{G}$ is linear over $id_{\bY_{\zet}}$;
we have $T\pi_{\zet}\circ\mathfrak{G} = id_{T\bX}\circ 
\bar{\pi}^{(r,k)}$, 
where $\bar{\pi}^{(r,k)}$ is the natural projection
$\bY_{\zet}\ucar{\bX} 
\cA^{(r,k)} \to T\bX$;
 for any pair $(\bar{\Lam},\bar{\Xi})$ $\in$
$\cA^{(r,k)}$, we have
$\mathfrak{G}([\bar{\Lam},\bar{\Xi}]) = [\mathfrak{G}(\bar{\Lam}), \mathfrak{G}(\bar{\Xi})]$.

\bDf
Let $\gam$ be a (local) section of $\bY_{\zet}$, $\bar{\Xi}$ 
$\in \cA^{(r,k)}$ and $\hat\Xi$ its gauge-natural lift. 
Following \cite{KMS93} we
define the {\em 
generalized Lie derivative} of $\gam$ along the vector field 
$\hat{\Xi}$ to be the (local) section $\pounds_{\bar{\Xi}} \gam : \bX \to V\bY_{\zet}$, 
given by
$\pounds_{\bar{\Xi}} \gam = T\gam \circ \xi - \hat{\Xi} \circ \gam$.\END
\eDf

The Lie derivative operator acting on sections of gauge-natural 
bundles is an homomorphism of Lie algebras; furthermore, for any vector field $\bar{\Xi} \in \cA^{(r,k)}$, the 
mapping $\gam \mto \pounds_{\bar{\Xi}}\gam$ 
is a first--order quasilinear differential operator and 
 for any local section $\gam$ of $\bY_{\zet}$, the mapping 
$\bar{\Xi} \mto \pounds_{\bar{\Xi}}\gam$ 
is a linear differential operator.
Moreover, we can regard $\pounds_{\bar{\Xi}}: J_{1}\bY_{\zet} \to V\bY_{\zet}$ 
as a morphism over the
basis $\bX$ and by using the canonical 
isomorphisms $VJ_{s}\bY_{\zet}\simeq J_{s}V\bY_{\zet}$ for all $s$, we have
$\pounds_{\bar{\Xi}}[j_{s}\gam] = j_{s} [\pounds_{\bar{\Xi}} \gam]$,
for any (local) section $\gam$ of $\bY_{\zet}$ and for any (local) 
vector field $\bar{\Xi}\in \cA^{(r,k)}$. 
We remark that, for any gauge-natural lift, the fundamental relation holds true:
$\hat{\Xi}_V\doteq (\mathfrak{G}(\bar{\Xi}))_{V}=- \pounds_{\bar{\Xi}}$.

\medskip

The splitting \eqref{jet connection} induces splittings in the spaces of forms
\cite{Vit98}; here and in the sequel we implicitly use
identifications between spaces of forms and spaces of bundle morphisms
which are standard in the calculus of variations (see, \eg
\cite{Kol83,KMS93,Kru73}).

 For $s \geq 0$, we consider the standard sheaves $\For^{p}_{s}$
of $p$--forms on $J_s\bY$.
  For $0 \leq q \leq s $, we consider the sheaves $\Hor^{p}_{(s,q)}$ and
$\Hor^{p}_{s}$ of {\em horizontal forms} with respect to the 
projections $\pi^s_q$ and $\pi^s_0$, respectively.
  For $0 \leq q < s$, we consider the subsheaves $\Con^{p}_{(s,q)}
\sub \Hor^{p}_{(s,q)}$ and $\Con^{p}{_s} \sub
\Con^{p}_{(s+1,s)}$ of {\em contact forms}, \ie horizontal forms 
valued into $\cC^{*}_{s}[\bY]$ (they have the property of vanishing 
along any section of the gauge-natural bundle). According to
\cite{Kru90,Vit98}, the fibered splitting
\eqref{jet connection} yields the {\em sheaf splitting}
$\Hor^{p}_{(s+1,s)}$ $=$ $\bigoplus_{t=0}^p$
$\Con^{p-t}_{(s+1,s)}$ $\wed\Hor^{t}_{s+1}$, which restricts to the inclusion
$\For^{p}_s$ $\sub$ $\bigoplus_{t=0}^{p}$
$\Con^{p-t}{_s}\wed\Hor^{t,}{_{s+1}^{h}}$,
where $\Hor^{p,}{_{s+1}^{h}}$ $\doteq$ $h(\For^{p}_s)$ for $0 < p\leq n$ and the map
$h$ is defined to be the restriction to $\For^{p}_{s}$ of the projection of
the above splitting onto the non--trivial summand with the highest
value of $t$.

Let 
$\eta\in\Con^{1}_{s}\wed\Con^{1}_{(s,0)}\wed\Hor^{n,}{_{s+1}^{h}}$;
then there is a unique morphism
$$
K_{\eta} \in \Con^{1}_{(2s,s)}\otimes\Con^{1}_{(2s,0)}\wed\Hor^{n,}{_{2s+1}^{h}}
$$
such that, for all $\Xi:\bY\to V\bY$,
$ C^{1}_{1} (j_{2s}\Xi\ten K_{\eta}) =E_{{j_{s}\Xi}\rfloor \eta}$,
where $C^1_1$ stands for tensor
contraction on the first factor and $\rfloor$ denotes inner product and $E_{{j_{s}\Xi}\rfloor \eta} =  (\pi^{2s+1}_{s+1})^* {j_{s}\Xi}\rfloor \eta+F_{{j_{s}\Xi}\rfloor \eta}$ (with $F_{{j_{s}\Xi}\rfloor \eta}$ a local divergence) is a uniquely defined global section of $\Con^{1}_{(2s,0)}\wed\Hor^{n,}{_{2s+1}^{h}}$ (see \cite{Vit98}). 

By an abuse of notation, let us denote by $d\ker h$ the sheaf
generated by the presheaf $d\ker h$ in the standard way.
We set $\Thd^{*}_{s}$ $\doteq$ $\ker h$ $+$
$d\ker h$. We have that 
$\diagramstyle[size=1.3em]
\begin{diagram}
0 &\rTo & \R_{Y} &\rTo & \Var^{*}_s
\end{diagram}$,
where $\Var^{*}_s=\For^{*}_s / \Thd^{*}_{s}$, is an exact resolution of the constant sheaf $ \R_{Y} $ \cite{Kru90}.
Consider the following   truncation

$\diagramstyle[size=1.3em]
\begin{diagram}
0 &\rTo & \R_{Y} &\rTo & \Var^{0}_s & \rTo^{\cE_0} &
\Var^{1}_{s} & \rTo^{\cE_{1}} & \dots  & \rTo^{\cE_{n}} &
\Var^{n+1}_{s}  & \rTo^{\cE_{n+1}} & \cE_{n+1}(\Var^{n+1}_{s})  
& \rTo^{\cE_{n+2}} & 
0,
\end{diagram}$

represented  by Vitolo in \cite{Vit98}.
A section $E_{d\lam}\doteq \cE_{n}(\lam) \in \Var^{n+1}_{s}$ is the {\em generalized higher 
order Euler--Lagrange type morphism} associated with $\lam$. 
The morphism $K_{\eta}$ previously introduced can be integrated by parts to provide a representation of the {\em generalized
Jacobi morphism} associated with $\lam$ \cite{PaWi03}, which then can be seen to be a generalized higher degree Euler--Lagrange type morphism.

\section{Jacobi equations and reductive split structure}\label{4}

Let $\lam$ be a Lagrangian and consider $\hat{\Xi}_{V}$ as a variation vector field. 
Let us set $\chi(\lam,\hat{\Xi}_{V})$ $\doteq$ 
 $C^{1}_{1} (\hat{\Xi}_{V}$ $\ten$ $K_{hd\cL_{j_{2s}\bar{\Xi}_{V}}\lam})$ $\equiv$  
$E_{j_{s}\hat{\Xi}\rfloor
hd\cL_{j_{2s+1}\bar{\Xi}_V}\lam}$. Because of linearity properties of
$K_{hd\cL_{j_{2s}\bar{\Xi}_{V}}\lam}$, and by using a global decomposition formula
due to Kol\'a\v{r}~\cite{Kol83}, we can decompose the morphism defined above as 
$\chi(\lam,\hat{\Xi}_{V})=E_{\chi(\lam,\hat{\Xi}_{V})}+ F_{\chi(\lam,\hat{\Xi}_{V})}$, 
where $F_{\chi(\lam,\hat{\Xi}_{V})}$ is a {\it local} horizontal differential which can be globalized by
fixing of a connection; however we do not fix any connection {\it a priori} here. 

\bDf
We call the morphism $\cJ(\lam,\hat{\Xi}_{V})$ $\doteq$
$E_{\chi(\lam,\hat{\Xi}_{V})}$  the {\em gauge-natural generalized Jacobi 
morphism} associated with the Lagrangian $\lam$ and the variation vector field
$\hat{\Xi}_{V}$. 
We call the morphism 
$\mathfrak{H}(\lam, \hat{\Xi}_{V})\doteq\hat{\Xi}_{V}\rfloor
\cE_{n}(\hat{\Xi}_{V}\rfloor\cE_{n}(\lam))$  the {\em gauge-natural
Hessian morphism} associated with $\lam$.\END 
\eDf
The morphism $\cJ(\lam,\hat{\Xi}_{V})$ is a 
{\em linear} morphism with respect to the projection 
$J_{4s}\bY_{\zet}\ucar{\bX}VJ_{4s}\cA^{(r,k)} \to J_{4s}\bY_{\zet}$.
Such a morphisms has been also represented on finite order variational sequence modulo
horizontal differentials \cite{FPV02} and thereby proved to be self-adjont along solutions of the
Euler--Lagrange equations, a result already well known for first order field theories \cite{GoSt73}. 
By resorting to the relation with the Hessian morphism \cite{PaWi07}, we shall prove here the same property in finite order variational sequences on gauge-natural bundles {\em without quotienting out horizontal differentials}.
As in the case of first order theories, we have in fact the following.
\bLm
The Hessian and thus the Jacobi morphism are symmetric self-adjoint morphisms.
\eLm
\bPf
Since $\del^{2}_{\mathfrak{G}}\lam \doteq  \cL_{\hat{\Xi}_{V}}\cL_{\hat{\Xi}_{V}}\lam =
\hat{\Xi}_{V}\rfloor \cE_{n}(\hat{\Xi}_{V}\rfloor\cE_{n}(\lam))$, 
we have $\mathfrak{H}(\lam, \hat{\Xi}_{V})= \del^{2}_{\mathfrak{G}}\lam$; 
furthermore,  being also $\del^{2}_{\mathfrak{G}}\lam = \cE_{n}(\hat{\Xi}_{V}\rfloor 
h(d\del\lam))$ (see the proof given in \cite{PaWi03}), then $\mathfrak{H}(\lam, \hat{\Xi}_{V})$ is self-adjoint.
Furthermore, we have 
\bEq\label{EQ}
\cJ(\lam,\hat{\Xi}_{V})\doteq E_{\chi(\lam,\hat{\Xi}_{V})} =\cE_{n}(\hat{\Xi}_{V}\rfloor 
h(d\del\lam)) =\mathfrak{H}(\lam, \hat{\Xi}_{V})\,. \QED
\eEq
\ePf

The Jacobi morphism $\cJ(\lam,\hat{\Xi}_{V})$ can be interpreted as an endomorphism of $J_{4s}V\cA^{(r,k)}$. In the following we concentrate on some geometric aspects of the space $\mathfrak{K}\doteq \ker\cJ(\lam,\hat{\Xi}_{V})$. Such a kernel defines generalized gauge-natural Jacobi equations \cite{PaWi03}, the solutions of which we call {\em generalized Jacobi vector fields}. It characterizes {\em canonical} covariant conserved quantities.
In fact, given $[\alp]\in \Var^{n}_{s}$, since the {\em variational Lie derivative} \cite{FPV98a} of classes of forms can be represented the variational sequence, we
have the corresponding version of the First Noether Theorem:
\bEq\label{noether I}
\cL_{j_{s}\Xi}[\alp] =
\ome(\lam,\hat{\Xi}_{V}) +
d_{H}(j_{2s}\hat{\Xi}_{V} \rfloor p_{d_{V}h(\alp)}+ \xi \rfloor h(\alp))\,,
\eEq
where we put  $\ome(\lam,\hat{\Xi}_{V}) \doteq \hat{\Xi}_{V}\rfloor\cE_{n}(\lam)  \doteq  -\pounds_{\bar{\Xi}} 
\rfloor \cE_{n} (\lam)$.

As usual,  $\lam$ is defined a
{\em gauge-natural invariant Lagrangian} if the gauge-natural lift
$(\hat{\Xi},\xi)$ of {\em any} vector
field $\bar{\Xi} \in \cA^{(r,k)}$ is a  symmetry for
$\lam$, \ie if $\cL_{j_{s+1}\bar{\Xi}}\,\lam = 0$. In this case, as an immediate consequence we have that
$\ome(\lam,\hat{\Xi}_{V})=
d_{H}(-j_{s}\pounds_{\bar{\Xi}}
\rfloor p_{d_{V}\lam}+ \xi \rfloor \lam)$.
The generalized {\em Bergmann--Bianchi morphism} \cite{Ber49} $\bet(\lam,\hat{\Xi}_{V})$ $\doteq$ $E_{\ome(\lam,\hat{\Xi}_{V})}$,
which  is nothing but the 
Euler--Lagrange morphism associated with the {\em new} Lagrangian 
$\ome(\lam,\hat{\Xi}_{V})$ defined on the fibered 
manifold $J_{2s}\bY_{\zet} \ucar{\bX} VJ_{2s}\cA^{(r,k)}\to \bX$. We proved that the generalized Bergmann-Bianchi morphism is canonically vanishing along $\mathfrak{K}$. This fact characterizes canonical covariant conserved Noether currents \cite{PaWi04}. 

Notice that since $\lam$  is  gauge-natural  invariant then 
$\cL_{j_{s+1}\hat{\Xi}}[\cL_{j_{s+1}\hat{\Xi}_{V}}\lam]$ $=$
 $\cL_{j_{s+1}[\hat{\Xi},\hat{\Xi}_{V}]}\lam+
 \cL_{j_{s+1}\hat{\Xi}_{V}}\cL_{j_{s+1}\hat{\Xi}}\lam$ $=$
 $\cL_{j_{s+1}[\hat{\Xi}_{H},\hat{\Xi}_{V}]}\lam$. However, we remark that the Lagrangian $\ome$ {\em is not gauge-natural invariant} unless, either
$[\hat{\Xi}_{H},\hat{\Xi}_{V}] $= $0$, or such a commutator is the gauge-natural lift of some infinitesimal principal automorphism. 

Nevertheless, along the kernel of the gauge-natural
generalized gauge-natural Jacobi morphism we have that 
$\cL_{j_{s+1}\bar{\Xi}_{H}}[\cL_{j_{s+1}\bar{\Xi}_{V}}\lam]\equiv 0$. 
Hence Bergmann--Bianchi identities are equivalent to the invariance condition $\cL_{j_{s+1}\bar{\Xi}}[\cL_{j_{s+1}\bar{\Xi}_{V}}\lam]$ $\equiv$  $0$ and can be suitably  interpreted as Noether identities associated with the invariance properties of the higher degree Euler--Lagrange morphism $\cE_{n}(\ome)$ \cite{PaWi08}.
As a consequence \cite{FPW05,FPW04} there exists a covariant $n$-form 
$\cH(\lam,\mathfrak{K})$ which can be interpreted as a Hamiltonian form for
$\ome(\lam,\mathfrak{K})$ on the Legendre bundle $\Pi\equiv 
V^{*}(J_{2s}\bY_{\zet} \ucar{\bX} VJ_{2s}\cA^{(r,k)})\wed (\owed{n-1}
T^{*}\bX)$.
Let then $\Ome$ be the multisimplectic form on the corresponding homogeneous Legendre bundle $\bZ\doteq T^{*}(J_{2s}(\bY_{\zet}\ucar{\bX}V\cA^{(r,k)}))\wed\For^{n-1}T^{*}\bX$.
Every Hamiltonian form $\cH$ admits a 
Hamiltonian connection $\gam_{\cH}(\lam,\mathfrak{K})$ such that the Hamilton equations 
$\gam_{\cH}(\lam,\mathfrak{K})\rfloor\Ome = d\cH(\lam,\mathfrak{K})$ hold true \cite{MaSa00}. In \cite{PaWi07} we proved that the Hamilton
equations for the
Hamiltonian connection form $\gam_{\cH}(\lam,\mathfrak{K})$ coincide 
with the kernel of the generalized gauge-natural Jacobi 
morphism. As a consequence $\mathfrak{K}$ is characterized as a vector subbundle \cite{FFPW08,FPW05,PaWi07,Win07}.
\bTh
The kernel $\mathfrak{K}$ defines a reductive structure on $W^{(r+4s,k+4s)}\bP$.
\eTh

\bPf
Being the Jacobi morphism self-adjoint its cokernel coincides with the cokernel of the adjoint morphism, thus we have that $\textstyle{dim}\mathfrak{K}=\textstyle{dim}\textstyle{Coker}\cJ$. If we further consider that $\mathfrak{K}$ is of constant rank because, as we just recalled, it is the kernel of a Hamiltonian operator, we are able to define the  split structure on $(VW^{(r+4s,k+4s)}\bP)$ $/$ $W^{(r+4s,k+4s)}_{n}\bG$, given by $\mathfrak{K}\oplus \textstyle{Im}\cJ$.

Let $\mathfrak{h}$ be the Lie algebra of right-invariant vector fields on $W^{(r+4s,k+4s)}\bP$ and $\mathfrak{k}$ the Lie subalgebra of generalized Jacobi vector fields defined as solutions of generalized Jacobi equations. 
 The Lie derivative of a solution of Euler--Lagrange equations  {\em with respect to a Jacobi vector field} is again a solution of Euler--Lagrange equations. However, the Lie derivative with respect to vertical parts of the commutator between the gauge-natural lift of a Jacobi vector field and (the vertical part of) a lift not lying in $\mathfrak{K}$  {\em is not} a solution of Euler--Lagrange equations.  Thus, since $\cJ$ is a projector and a derivation of $\mathfrak{h}$, it is easy to see that the split structure is also reductive, being $[ \mathfrak{k},\textstyle{Im}\cJ]=\textstyle{Im}\cJ$.
\QED\ePf

\bRm
As a consequence generalized Jacobi vector fields define a kind of reductive gauge-natural lift: we are concerned with the reduction of the structure bundle $W^{(r,k)}\bP$ to a subbundle with structure group the subgroup of a differential group of the base manifold; thus recovering the geometric framework of reductive G-structures, reductive lifts and induced reductive Lie derivatives of sections, as  constructed in \cite{GoMa03}.\END
\eRm

\medskip

Our investigations are mainly concerned with the existence of covariant canonically defined conserved currents \cite{PaWi04}. More precisely, we consider Lagrangian field theories which are assumed to be invariant with respect to the action of a gauge-natural group $W^{(r,k)}_{n}\bG$ defined as the semidirect product of a $k$-th order differential group of the base manifold with the group of $r$-th order velocities in $\bG$ (see Section \ref{1}). In fact, notice that the group $\Diff (\bX)$ {\em is not} canonically embedded into $\textstyle{Aut}(\bP)$ (see, in particular, the discussion of this aspect presented in \cite{FFR03,GoMa03,Mat03}). We denote by $\textstyle{Aut}(\bP)$ the group of {\em all} automorphisms of the underling principal bundle $\bP$, not the ``gauge group'' of {\em vertical} principal automorphisms, as it is sometimes  done in Physics.  In other words, we are faced with the following general problem: we know how fields transform corresponding to a transformation in $\bP$ but we do not know how fields transform under changes of coordinates in the base manifold, so that Lie derivatives with respect to infinitesimal base transformations cannot be defined neither in a natural nor in a canonical way, at least {\em a priori}. 

It is a well known fact that the covariance of the Lagrangian and thus of the Euler-Lagrange equations does not guarantee the corresponding covariance of Noether conserved quantities (a well known example is the Einstein energy-momentum pseudotensor which was covariantized by Komar {\em via} the introduction of a connection). In all generality, it is a well known fact that one need the fixing of a linear connection on the base manifold and of a principal connection on the principal bundle to get covariant conserved quantities in gauge-natural field theories (this is the outcome of the fact that a global Poincar\'e--Cartan form can be defined only by fixing such a couple of connections \cite{FaFr03}). However, we showed that a {\em canonical} determination of Noether conserved quantities is always possible on a reduced bundle of $W^{(r,k)}\bP$ {\em completely determined by the original  $W^{(r,k)}_{n}\bG$-invariant variational problem}, without fixing any connection {\em a priori}. Instead connections can be characterized by means of such canonical reduction \cite{FFPW08}.

Concerning this last point, we stress that several aspects of the geometric formulation of field theories on bundles associated with principal bundles, with different techniques and adopting alternative points of views, for example stressing the {\em r\^ole} of the Poincar\'e--Cartan form, or formulated on infinite order jet prolongations, have been the subject of a widespread research activity since the {\em Seventies} of last Century. Among them, of relevant interest for Physics, the study of reductions of the underlying principal bundle {\em \ie reductions of the  gauge group of the theory}. We refer in particular to \cite{BGMS05a,BGMS05b,CaGaRa01,CaRa03,CaMa03,GoMa03,Ja06,Ja07,Sar06}. Most of such researches were essentially motivated by possible generalizations of the Utiyama Theorem \cite{Uti56}.  

The papers \cite{CaGaRa01,CaRa03,CaMa03} are mainly concerned with the reduction of a given variational problem on principal bundles. The word `reduction' here takes a strictly variational meaning: one considers reduced variational problems, given by the reduced Lagrangian on $J_{1}\bP/\bG$, describing so--called Euler--Poincar\'e equations.  {\em Assuming the existence of a reduction of the principal bundle $\bP$}, the corresponding reduced variational problem describing Lagrange--Poincar\'e equations is characterized and the reconstruction of the original variational problem from the reduced one is expressed by means of a certain condition on the curvature of a relevant suitably constructed principal connection. The `semidirect product reduction' considered in \cite{CaRa03} is still of purely gauge nature, and does not involve the semidirect action of a differential group of the base manifold on the group of $r$-th order velocities in $\bG$.
The papers \cite{BGMS05a,BGMS05b,Sar06} are rather closer to our approach: the {\em existence}  of  reduced gauge transformations {\em induced} by the variational problem itself (in particular by the existence of Noether identities) is investigated; however only purely gauge theories are studied and the problem of the relation of gauge transformations with diffeomorphisms of the base manifold is not considered. 
Utiyama-like theorems in the case of principal bundles having the structure of a gauge-natural prolongation \cite{Ja06,Ja07} are mainly purely geometric constructions concerning naturality and functorial aspects, without direct relation with the calculus of variations.

We stress again that the main point at the base of our paper \cite{PaWi03} was the possibility of relating the vertical components of a gauge natural lifts with the horizontal ones: \ie relate gauge infinitesimal transformations with infinitesimal transformations of the base manifold, a first step towards the theory of a unified field.  Bundles of fields associated with the class of principal bundles obtained as gauge-natural prolongations of principal bundles \cite{KMS93} have a {\em richer} structure than principal bundles {\em tout court}.  In fact a relation between gauge charges and energy-momentum-like conserved quantities can be obtained when considering invariant variational problems on such a subclass of principal bundles.

As a matter of example let us in fact consider the Lie derivative of spinor fields. It is possible to recover the Kosmann lift (which is not a natural lift) in terms of the reductive lift induced by the kernel of the generalized gauge-natural Jacobi morphism.  The Jacobi equations for the well known Einstein--(Cartan)--Dirac Lagrangian just implies that -- if   $\bar{\Xi}^{a}_{v\,b}=\bar{\Xi}^{a}_{b} -\ome^{a}_{b\mu}\xi^{\mu}$ is the
vertical part of $\bar{\Xi}$ with respect to the spin connection $\ome$ and  the corresponding superpotential for the Noether current is given by $\nu(\lam,\bar{\Xi})\doteq -\frac{1}{2k}\bar{\Xi}_{v}^{ab}\eps_{ab}$ -- then we obtain $\bar{\Xi}_{v}^{ab}=-\tilde{\nabla}^{[a}\xi^{b]}$, \ie the  well known Kosmann lift, where
$\tilde{\nabla}$ is the covariant derivative with respect to the standard transposed connection on
the bundle of spin-tetrads. 
On the other hand, the Lie derivative of spinor fields can be expressed 
in terms of  the horizontal part of $\hat{\Xi}$ {\em with respect to 
the spinor-connection} $\hat{\Xi}_{h}$: consequently  we obtain a constraint on the corresponding connection 
$\tilde{\ome}$ on the spinor bundle, as well as on the superpotential corresponding to $\bar{\Xi}_{v}^{ab}$. In this way we characterize a unique canonical superpotential invariant with respect to the reduced group. Here we used a quite standard notation \cite{FaFr03}; for further details we refer to \cite{FFPW08,PaWi08,Win07}. 

\subsection*{Acknowledgments}
The authors are grateful to Prof. I. Kol\'a\v{r} for useful discussions and thank the unknown referees for suggestions which improved the presentation. 
This research has been supported by MIUR--PRIN(2005), the University of Torino and by an E\v{C}C grant (E.W.).


\end{document}